\documentclass[aps,twocolumn,floatfix,superscriptaddress,nofootinbib]{revtex4-2}
\usepackage[english]{babel}
\usepackage{booktabs}
\usepackage{graphicx}% Include figure files
\usepackage{braket}
\usepackage{stix}
\usepackage{xcolor}
\usepackage{amsfonts} 
\usepackage{appendix}
\usepackage{pifont}
\usepackage{amsmath}
\usepackage{multirow}
\usepackage{dcolumn}% Align table columns on the decimal point
\usepackage{soul}
\usepackage{bm}% bold math
\usepackage{amsthm}
\usepackage{enumitem}
\usepackage{amsmath}

\usepackage{xr-hyper}
\usepackage{hyperref}
\hypersetup{colorlinks,
	citecolor=blue
}
\usepackage{cleveref}
\usepackage{apptools}
\usepackage{svg}
\usepackage{tikz}
\usetikzlibrary{quantikz2}
\AtAppendix{\counterwithin{lemma}{section}}

\makeatletter
\newcommand*{\addFileDependency}[1]{% argument=file name and extension
  \typeout{(#1)}
  \@addtofilelist{#1}
  \IfFileExists{#1}{}{\typeout{No file #1.}}
}
\makeatother

\usepackage{fancyhdr,graphicx,amsmath,amssymb}

\usepackage{graphicx}
\usepackage{afterpage}
\usepackage[ruled,vlined]{algorithm2e}
\include{pythonlisting}

\begin{document}

\preprint{APS/123-QED}

\title{Superdiffusion resilience in Heisenberg Chains with 2D interactions on a quantum processor}% Force line breaks with \\

\author{Keerthi Kumaran$^\bigstar$}
\affiliation{Department of Physics and Astronomy, Purdue University, West Lafayette, IN 47907, USA}
\affiliation{Quantum Science Center, Oak Ridge National Laboratory, Oak Ridge, TN 37831, USA}

\author{Manas Sajjan$^{\bigstar,\diamondsuit  }$}
\email{msajjan@ncsu.edu}
\affiliation{Quantum Science Center, Oak Ridge National Laboratory, Oak Ridge, TN 37831, USA}
\affiliation{Department of Chemistry, Purdue University, West Lafayette, IN 47907, USA}
\affiliation{Department of Electrical and Computer Engineering, North
Carolina State University, Raleigh, NC 27606, USA}

\author{Bibek Pokharel$^{\,\diamondsuit}$}
\email{bibek.pokharel@ibm.com}
\affiliation{IBM Quantum, IBM T.J. Watson Research Center, Yorktown Heights, New York 10598, USA}

\author{Kevin Wang}
\affiliation{Department of Physics, University of California, Berkeley, California 94720, USA}

\author{Joe Gibbs}
\affiliation{School of Mathematics and Physics, University of Surrey, Guildford, GU2 7XH, UK}
\affiliation{AWE, Aldermaston, Reading, RG7 4PR, UK}

\author{Jeffrey Cohn}
\affiliation{IBM Quantum, Almaden Research Center, San Jose, CA, 95120 USA}

\author{Barbara Jones$^{\, \spadesuit}$}
\affiliation{IBM Quantum, Almaden Research Center, San Jose, CA, 95120 USA}

\author{Sarah Mostame$^\diamondsuit $}
\email{sarah.mostame@ibm.com}
\affiliation{IBM Quantum, IBM T.J. Watson Research Center, Yorktown Heights, New York 10598, USA}

\author{Sabre Kais}
\affiliation{Quantum Science Center, Oak Ridge National Laboratory, Oak Ridge, TN 37831, USA}
\affiliation{Department of Chemistry, Purdue University, West Lafayette, IN 47907, USA}
\affiliation{Department of Electrical and Computer Engineering, North
Carolina State University, Raleigh, NC 27606, USA}

\author{Arnab Banerjee$^\diamondsuit $}
\email{arnabb@purdue.edu}
\affiliation{Department of Physics and Astronomy, Purdue University, West Lafayette, IN 47907, USA}
\affiliation{Quantum Science Center, Oak Ridge National Laboratory, Oak Ridge, TN 37831, USA}

%%%%%%%%%%%%%%%%%%%%%%%%%%%%%%%%%%%%%%
\def\thefootnote{$^\bigstar$}\footnotetext{ These authors contributed equally to this work.}

\def\thefootnote{$^\spadesuit$}\footnotetext{Deceased author. We note with sadness that our co-author, Dr. Barbara Jones, passed away during the preparation of this manuscript. We are grateful for her invaluable contributions to this work.}

\def\thefootnote{$^\diamondsuit $}\footnotetext{Corresponding authors }

%%%%%%%%%%%%%%%%%%%%%%%%%%%%%%%%%%%%%%

\begin{abstract}
Observing superdiffusive scaling in the spin transport of the integrable 1D Heisenberg model is one of the key discoveries in non-equilibrium quantum many-body physics. Despite this remarkable theoretical development and the subsequent experimental observation of the phenomena in KCuF$_3$, real materials are often imperfect and contain integrability breaking interactions. Understanding the effect of such terms on the superdiffusion is crucial in identifying connections to such materials. Current quantum hardware has already ascertained its utility in studying such non-equilibrium phenomena by simulating the superdiffusion of the 1D Heisenberg model. In this work, we perform a quantum simulation of the superdiffusion breakdown by generalizing the superdiffusive Floquet-type 1D Heisenberg model to a general 2D model. We comprehensively study the effect of different 2D interactions on the superdiffusion breakdown by tuning up their strength from zero, corresponding to the 1D Heisenberg chain, to finite nonzero values. We observe that certain 2D interactions are more resilient against superdiffusion breakdown than others and that the $SU(2)$ preserving 2D interaction has the highest resilience among all the 2D interactions we study. Importantly, this observed resilience has direct implications for sustaining superdiffusive spin transport in two-dimensional lattices. We reason out the relative resilience against the superdiffusion breakdown through an analysis of the scattering coefficients off the 2D interaction in otherwise 1D chains. The relative resilience of different interaction types against superdiffusion breakdown was also captured in quantum hardware with remarkable accuracy, further establishing the current quantum hardware's applicability in simulating interesting non-equilibrium quantum many-body phenomena.
\end{abstract}

\maketitle

\section{Introduction}

Quantum many-body (QMB) simulation is one of the most promising areas for demonstrating quantum advantage \cite{QSim-Advantage, Qsim_1}. Digital quantum simulators with fast developing \cite{QEC} fault-tolerant hardware could open a wide range of applications.  However, even before fault tolerance, quantum simulators are already being used to solve utilitarian QMB problems  \cite{utility,utility_2,utility3,utility4,utility5,utility6}. Understanding non-equilibrium dynamics \cite{non_equi_book} of quantum systems is an active area of research in the QMB physics and unsurprisingly interests the quantum simulation community as well  (see\cite{non_equi_1,non_equi_2, lee2025digitalquantumsimulationspin,non_equi_3}).

Quantum spin systems, while governed microscopically by the Schrödinger equation, exhibit emergent coarse-grained hydrodynamic behavior shaped by their conservation laws. In one dimension, integrable quantum systems are typically expected to display ballistic transport ($\propto t^{-1}$) due to their extensive set of conserved quantities, especially in non-interacting or trivially integrable models. However, this expectation can break down in interacting integrable systems. For example, in the anisotropic Heisenberg (XXZ) spin-1/2 chain, transport behavior depends sensitively on the anisotropy parameter ($\Delta$): ballistic transport occurs for $\Delta < 1$, diffusive transport ($\propto t^{-1/2}$) for $\Delta > 1$, and superdiffusive spin transport ($\propto t^{-2/3}$) emerges at the isotropic point $\Delta = 1$, where the dynamics follow a diffusion equation in nonlinearly scaled time, challenging conventional hydrodynamic predictions~\cite{Znidaric2011,integrable_1,integrable_4,integrable_5,integrable_6,integrable_7,integrable_9}.

The first theoretical characterization of diffusion in integrable systems was presented in~\cite{DeNardis2018Diffusion}, where generalized hydrodynamics was extended to include Navier-Stokes-type corrections. More recently, a phenomenological mapping from the Heisenberg chain to the Kardar-Parisi-Zhang (KPZ) equation was proposed in~\cite{Ware2023KPZ}, providing a nonlinear fluctuating hydrodynamic theory that explains the emergence of KPZ scaling in isotropic spin chains. At this SU(2)-symmetric point, infinite-temperature spin-spin correlations exhibit anomalous scaling $\propto t^{-2/3}$, characteristic of the KPZ universality class, previously known only in classical systems~\cite{KPZ_classical_1, KPZ_classical_2, KPZ_classical_3} and certain stochastic quantum settings~\cite{KPZ_quantum_1, KPZ_quantum_2, KPZ_quantum_3}. This result has since been extended theoretically to low-temperature regimes~\cite{JM_delayed_onsets}, demonstrated experimentally~\cite{KPZ_exp_1, KPZ_exp_2, Joshi_2022}, and reproduced in quantum hardware simulations~\cite{Keenan_2023, Elliot}, highlighting the utility of quantum devices in probing emergent nonequilibrium behavior in spin systems.

\begin{figure*}
    \centering
    \includegraphics[width=1\textwidth]{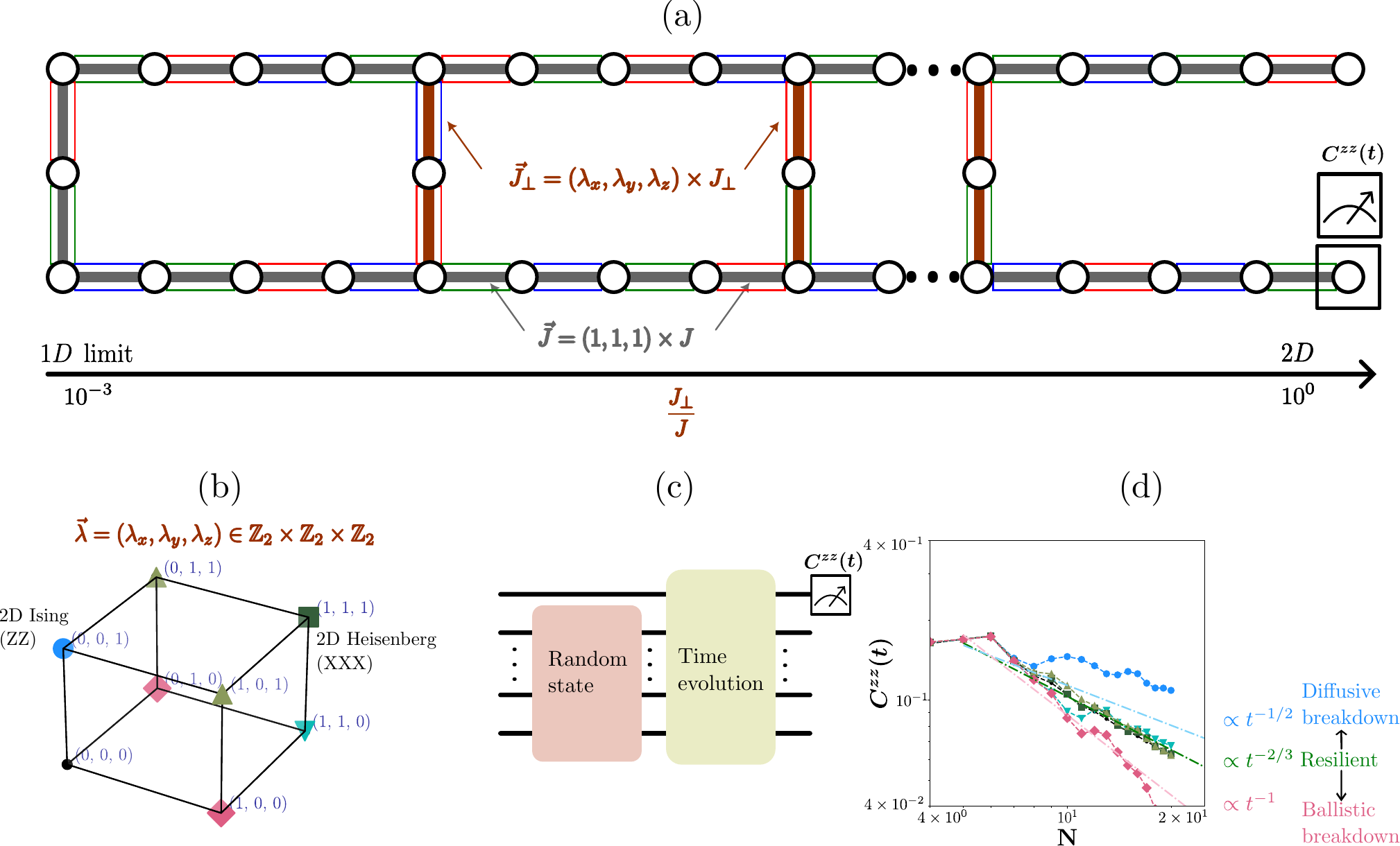}
    \caption{\textbf{Summary:} \textbf{(a)} We study the extension of the 1D Floquet Heisenberg Hamiltonian, represented by black bonds, known to exhibit superdiffusive scaling in discrete time evolution \cite{Discrete_1D_1, KPZ_floquet}, to a 2D Floquet model native to the heavy-hex lattice, represented by the vertical brown bonds. Discrete time evolution is performed by sequentially applying three types of bonds (colored red, green, and blue) at a kicking period $\tau$, with interaction strengths tuned relative to the 1D Heisenberg interaction and expressed as $\vec{J_\perp} = \vec{\lambda} \times J_\perp$ using a representative set of vectors \textbf{(b)} $\vec{\lambda}$ defined in Eq.~\ref{interaction_type}. \textbf{(c)} We measure the infinite temperature auto spin-spin correlation function of the edge probe qubit marked by the square box using a quantum algorithm \cite{Richter_Pal} to study \textbf{(d) }the breakdown and resilience of superdiffusion under different 2D interaction types.} 
    \label{fig:Scheme}
\end{figure*}
Materials in nature often contain a few terms that break integrability, which can cause any initial superdiffusive behavior to transition into diffusive behavior at late times. Hence, understanding the onset of superdiffusion breakdown is crucial for connecting theoretical models to real materials. The effect of integrability-breaking terms in classical integrable models has been explored in various works~\cite{HIS_chain,breaking_classical_1,breaking_classical_2}. Following this, the stability of superdiffusion under perturbations has also been studied in quantum systems ~\cite{DeNardis2021Superdiffusion}. Ref. ~\cite{breaking_quantum} recently discussed this effect in a quantum ladder model inspired by quasi-1D KCuF$_3$, showing that $SU(2) \times SU(2)$ symmetry-preserving perturbations to two unperturbed Heisenberg chains lead to a slower onset of superdiffusion breakdown compared to symmetry-breaking perturbations. Related analog quantum simulations of Heisenberg spin ladders~\cite{analog_breaking} and quantum gas microscopy experiments~\cite{Qgas} have reported deviations toward diffusive behavior. Similarly, deviations toward ballistic behavior were observed in a non-integrable setting~\cite{breaking_rydberg}.

In this work, we propose a digital quantum simulation of a 2D model shown in Fig.~\ref{fig:Scheme}, embeddable in a heavy-hex quantum hardware topology. This model converges to a Floquet~\cite{floquet_gen} 1D Heisenberg chain, known to exhibit superdiffusive behavior~\cite{Discrete_1D_1, KPZ_floquet}, when the 2D interaction strengths are tuned down to zero. Tuning up the 2D interaction allows us to study their effect on superdiffusion of the 1D model. We provide a comprehensive analysis on superdiffusion breakdown by studying 2D interactions of various interaction types, correlation measurement directions, and the location(s) of the 2D interactions/rungs with respect to the probe qubit whose spin transport we wish to study. Our results show that only the 2D interaction type affects the nature of superdiffusion breakdown, and the 2D interaction type preserving the SU(2) symmetry shows higher resilience against superdiffusion breakdown than other interaction types. Characterizing the resilience of superdiffusion under various 2D interactions offers insight into how such transport phenomena might be preserved in realistic two-dimensional systems. Within the interaction types that break the superdiffusion, we observe breakdown towards both diffusive and ballistic regimes with varying degrees of resilience within each of these classes. We reason out the relative resilience against the superdiffusion breakdown through an analysis of the scattering coefficients off the 2D interaction in otherwise 1D chains. We observe the relative resilience of interaction types in our hardware simulations on a minimal system size with significant accuracy, further bolstering the current hardware's capability in studying the non-equilibrium quantum spin transport behavior.

\section{Description of the model}
\label{model_description}
To study superdiffusion breakdown in the 1D Heisenberg (XXX) chain at infinite temperature due to 2D interaction terms,  we introduce a 2D model of a spin chain that converges to a 1D XXX chain when the 2D interaction strengths approach zero (see Fig.\ref{fig:Scheme}).

It is known that a Floquet Heisenberg Hamiltonian of the form in Eq.~\ref{1D_model} shows KPZ-like scaling~\cite{Discrete_1D_1,KPZ_floquet}.

\begin{equation}
    H_{1D}^\tau(t) = H_{\text{even-bonds}} + \tau H_{\text{odd-bonds}} \sum_{N \in \mathbb{Z}} \delta(N\tau - t)
    \label{1D_model}
\end{equation}

Let the spin sites be arranged linearly and labeled from 1 to \( n\). In Eq.~\ref{1D_model}, the Hamiltonians \( H_{\text{even-bonds}} \) and \( H_{\text{odd-bonds}} \) represent sets of nearest-neighbor XXX interactions acting on alternating adjacent pairs of spin sites. Specifically, \( H_{\text{even-bonds}} \) acts on site pairs such as  (2,3), (4,5), etc., while \( H_{\text{odd-bonds}} \) acts on the complementary set of pairs like (1,2), (3,4), (5,6), etc. Together, they span all adjacent spin pairs in the system. The second term on the RHS applies \( H_{\text{odd-bonds}} \) periodically at a dimensionless kicking interval \( \tau \). This discrete model offers an advantage over its continuous counterpart, as the superdiffusive scaling can be simulated on a quantum computer using a simple first-order Trotterization, applying \( H_{\text{even-bonds}} \) followed by \( H_{\text{odd-bonds}} \), as shown in Eq.~\ref{1D_evol} and demonstrated in~\cite{Keenan_2023}.

 \begin{equation}
     e^{-iH_{1D} (n\tau)} =  (e^{-i H_{\text{even-bonds}} \tau}e^{-iH_{\text{odd-bonds}} \tau})^n
     \label{1D_evol}
 \end{equation}
 
In our 2D model, we consider the Hamiltonian in Eq.~\ref{2D_model}.

\begin{equation}
    H_{2D}(t) = H_r +  \lim_{\epsilon \rightarrow 0^+} \tau H_g \sum_{N \in \mathbb{Z}} \delta(N\tau - \epsilon - t) + \tau H_b \sum_{N \in \mathbb{Z}} \delta(N\tau - t)
    \label{2D_model}
\end{equation}

In Eq.~\ref{2D_model}, the Hamiltonians \( H_i \) with \( i \in \{r, g, b\} \) correspond to sets of nearest-neighbor XXX interactions acting on distinct bond types labeled in Fig.~\ref{fig:Scheme} using red, green, and blue boxes, respectively. The term \( \epsilon \rightarrow 0^+ \) ensures that the exact time evolution of this model translates to performing a first-order Trotterization, which is natively compatible with a heavy-hex lattice, and is given by:

\begin{equation}
    e^{-iH_{2D}(N\tau)} = \left(e^{-iH_r \tau} e^{-iH_g \tau} e^{-iH_b \tau}\right)^N
    \label{2D_evol}
\end{equation}

 In this model, we consider two-local interactions $h_{i,j}$ composed of the terms $\sigma_x^{i}\sigma_x^{j},\sigma_y^{i}\sigma_y^{j}$ and $\sigma_z^{i}\sigma_z^{j}$ with strengths $\frac{\lambda_x}{4},\frac{\lambda_y}{4}$ and $\frac{\lambda_z}{4}$,  respectively. Let us denote such a two-local Hamiltonian $h_{i,j}$  by the vector $\vec{\lambda} = (\lambda_x,\lambda_y,\lambda_z) $, so that
\begin{equation}
\label{interaction_type}
   \vec{\lambda} \equiv  \frac{\lambda_x}{4} \sigma_x^i\sigma_x^j + \frac{\lambda_y}{4}\sigma_y^i\sigma_y^j + \frac{\lambda_z}{4}\sigma_z^i\sigma_z^j  = h_{i,j}
\end{equation}
With this notation, 1D XXX chain with interaction strength $J$ is  $\vec{J} = (1,1,1) \times J$. To have a minimalistic but versatile discussion on the superdiffusion breakdown due to 2D interactions, we focus only on the 2D interactions of strength $J_\perp$ and belong to the interaction types of the form shown in Eq \ref{int_types_1} and \ref{int_types_2}
 
 \begin{eqnarray}
 \label{int_types_1}
 \vec{J}_{\perp} = (\lambda_x,\lambda_y,\lambda_z) \times J_\perp = \vec{\lambda}  \times J_\perp \\
 \label{int_types_2}
 (\lambda_x,\lambda_y,\lambda_z) \in \mathbb{Z}_2 \times \mathbb{Z}_2 \times \mathbb{Z}_2 \\
 (J_\perp,J) \in \mathbb{R}\times\mathbb{R}
 \end{eqnarray}
 The time evolution in Eq.\ref{2D_evol} converges to the RHS of Eq. \ref{1D_evol} when $J_\perp / J \rightarrow 0$.  Hence, the 2D interactions in our model can be thought of as a perturbation to the 1D model and their effect on the superdiffusive scaling can be naturally simulated in any quantum hardware with heavy-hex topology such as IBM Quantum's hardware.

 In addition to $\vec{J_\perp}/J$, the model we describe has an inherent parameter $\tau$. By setting $\tau << 1$ while keeping the overall time of evolution fixed, we can simulate the superdiffusion breakdown of the continuous-time Hamiltonian $H_r + H_g + H_b$. Understanding continuous-time Hamiltonians is more interesting than the Floquet-type Hamiltonians owing to their relevancy in simulating real materials \cite{KPZ_exp_1}. However, setting a lower $\tau$ means we need to evolve our system (Eq.\ref{2D_evol}) with a higher number of trotter steps to see the superdiffusive scaling and its breakdown. This is challenging to simulate as it would result in longer depths in the quantum circuits and hence higher susceptibility to noise. And, $\tau >> 1$ is far away from the continuous-time model and is subject to a quick equilibration in its time evolution as discussed in Section \ref{tau_optimisation}. It was also recently shown to behave differently compared to continuous time model \cite{trotterized_applications_3}.  As a result, we choose $\tau = 1$ so that the number of trotter steps is $\sim$ 20 but simultaneously replicate the continuous-time model's behavior as shown in Fig \ref{dt_optimisation}.

\section{Results and Discussion}
\begin{figure*}
    \centering
    \includegraphics[width=1\linewidth]{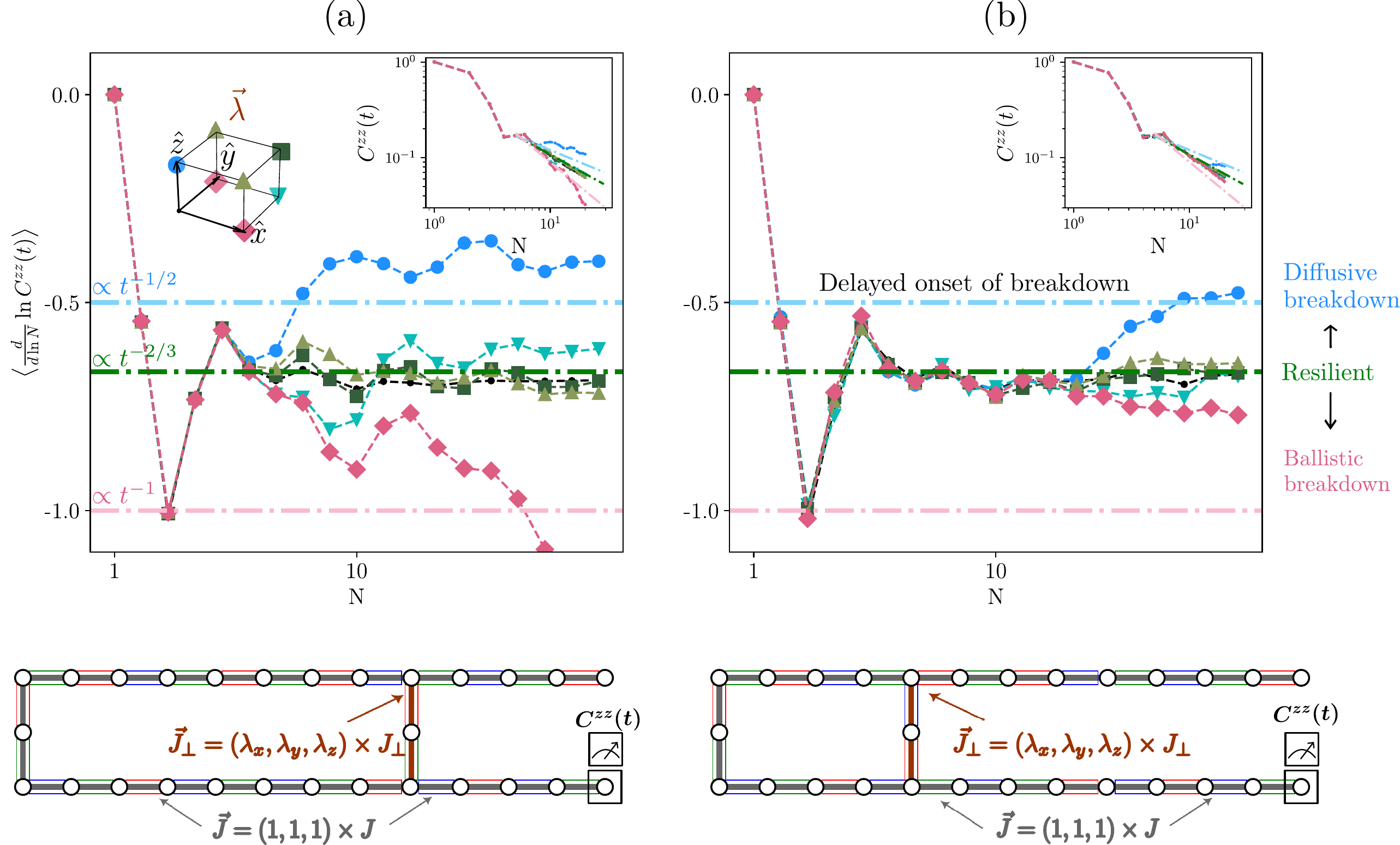}
      \caption{\textbf{Different types of superdiffusion breakdown} Using noiseless simulations of a 28-qubit system based on the model in Fig.~\ref{model_description}, we show that superdiffusion ($\propto t^{-2/3}$) can break down toward either the diffusive ($\propto t^{-1/2}$) or ballistic ($\propto t^{-1}$) regimes, depending on the interaction type. The figure presents running averages of scaling exponents $\left\langle \frac{d(\ln(C^{zz}(t)))}{d(\ln(N))} \right\rangle$, compared against the 1D reference model ($\vec{\lambda} = (0,0,0)$), with insets showing the corresponding $C^{zz}(t)$. Notably, $\vec{\lambda} = (1,1,1)$ remains most resilient against breakdown. Diffusive breakdown is observed for $\vec{\lambda} = (0,0,1)$ and $(1,1,0)$, with the latter showing greater resilience, while ballistic breakdown occurs for $\vec{\lambda} = (1,0,0)$ and $(1,0,1)$, with the latter being more resilient. All simulations were performed with a fixed interaction strength ratio of $\frac{J_{\perp}}{J} = 1.0$, by probing the 2D rung either closer (a) or farther (b) from the activated probe qubit; probing the farther qubit delays, but does not alter, the nature of the breakdown.
}
  
    \label{interaction_type_influence}
\end{figure*}

In our study, we investigate the breakdown of superdiffusive scaling by analyzing the spin-spin autocorrelation function at infinite temperature. The central observable is the autocorrelation function of the probe qubit, defined as
\begin{equation}
C_{pp}^{ii}(t) = \langle \sigma_p^i(0)\sigma_p^i(t) \rangle = \frac{1}{2^n} \text{Tr} \left[ \sigma_p^i(0)\sigma_p^i(t) \right],
\end{equation}
where $p$ denotes the position of the probe qubit in the chain (see Fig.~\ref{fig:Scheme}), $\sigma^i$ with $i \in \{x, y, z\}$ are Pauli operators corresponding to spin components, and $n$ is the total number of spins in the system. These autocorrelation functions are highly sensitive probes of spin transport, capturing the temporal evolution of local spin excitations.

This formulation is well suited for studying emergent hydrodynamic behavior in quantum many-body systems. In particular, the scaling behavior of $C_{pp}^{ii}(t)$ reveals the nature of transport, whether ballistic, diffusive, or superdiffusive, and provides direct insight into the breakdown of any anomalous spin transport under perturbations. Through linear response theory~\cite{KPZ_original, Kubo_formulas}, these correlations are intimately connected to spin currents across slightly polarized domain walls, allowing us to probe the mechanisms that govern transport and its breakdown.

For a given direction of correlation measurement, $i$,  we have $7$ different interaction types and hence there are $3 \times 7 = 21$ different sets of experiments to be run for any interaction strength $\frac{{J}_\perp}{J}$ to fully characterize the model. However, owing to the permutational symmetry of the Pauli operators in our model, we have several equivalencies in these experiments. Let the experiment measuring $C_{pp}^{ii}(t)$ for a 2D interaction of the type $(\lambda_x,\lambda_y,\lambda_z)$ be denoted by $\langle \lambda_x,\lambda_y,\lambda_z \rangle_{ii}$. Then we have the following equivalencies (Eqs \ref{equivalencies_1} and \ref{equivalencies_2})\\
\begin{eqnarray}
    \label{equivalencies_1}
    \langle \lambda_x,\lambda_y,\lambda_z \rangle_{zz} \equiv \langle \lambda_z,\lambda_y,\lambda_x\rangle_{xx} \equiv \langle \lambda_x,\lambda_z,\lambda_y\rangle_{yy} \\    \label{equivalencies_2}
    \langle 1,0,\lambda_z\rangle_{zz} \equiv \langle 0,1,\lambda_z\rangle_{zz} 
\end{eqnarray}\\
With the help of these equivalencies, for the rest of the work, we focus on just the $zz$ correlations of the interaction types 
\begin{equation}
   \label{interaction_type_compre}
    \vec{\lambda} = (0,0,1), (1,0,0), (1,1,0), (1,0,1), \text{and } (1,1,1)
\end{equation}

We follow the quantum algorithm described in \cite{Richter_Pal} to compute the $zz$ autocorrelation functions of the probe qubit. This algorithm requires a random state preparation with $c$ cycles of random two-qubit unitaries applied to $r$, $g$ and $b$-type bonds in that order and first-order trotterization with a time-step length of $\tau$. Refer to Appendix \ref{implementation_details} for details on the implementation.  Although we measure only $zz$ correlations, any general infinite-temperature correlation function $C^{nn}_{pp}(t)$ along a direction $\hat{n} = (n_x, n_y, n_z)$ can be reconstructed using
\[
C^{nn}_{pp}(t) = n_x^2 C^{xx}_{pp}(t) + n_y^2 C^{yy}_{pp}(t) + n_z^2 C^{zz}_{pp}(t).
\]
Each $C^{ii}_{pp}(t)$ can be mapped to a $C^{zz}_{pp}(t)$ measurement under a corresponding interaction type, as defined in Eq.~\ref{interaction_type_compre}.  
\begin{figure*}
    \centering
    \includegraphics[width=0.9\textwidth]{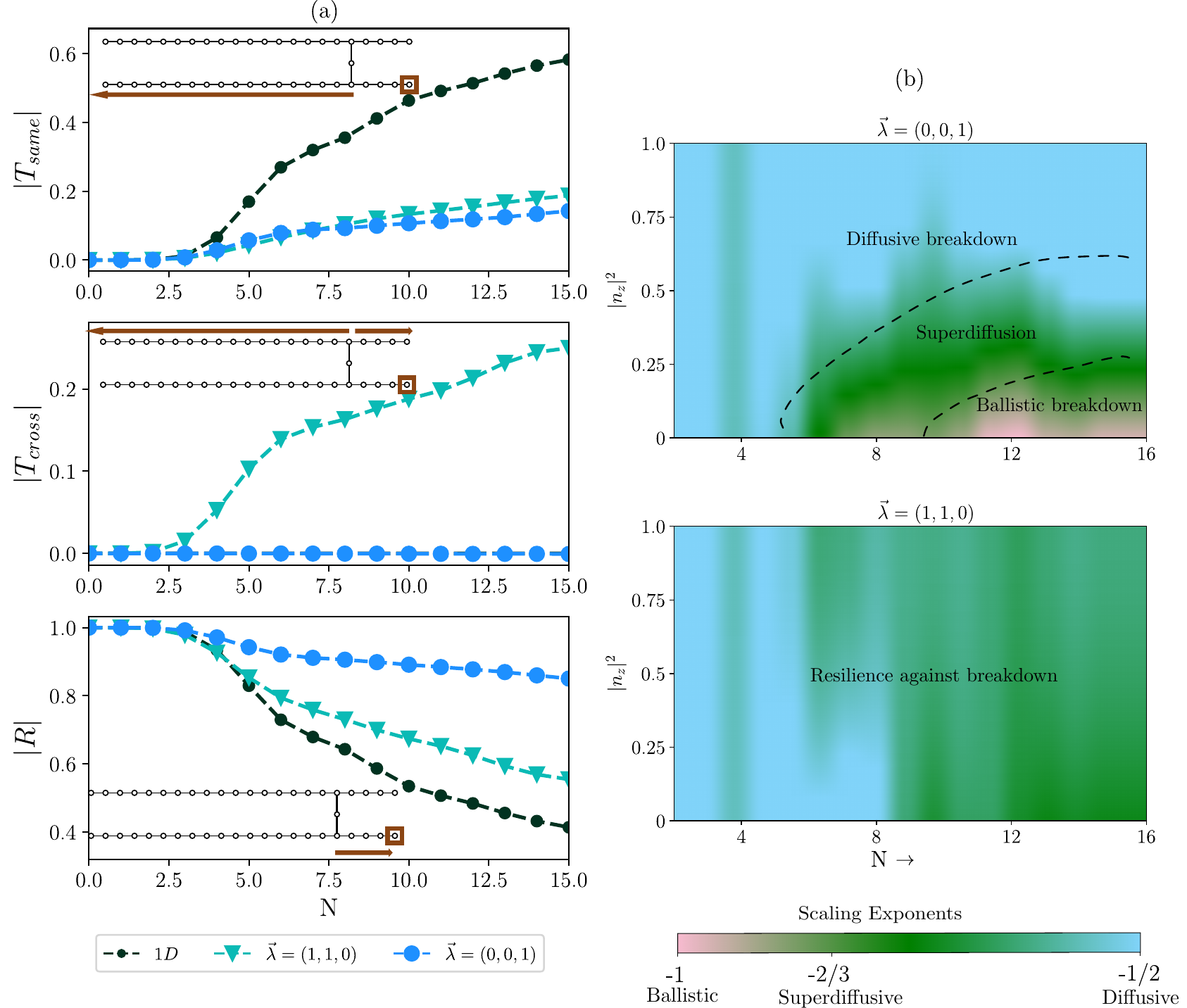}
    \caption{\textbf{(a) Scattering coefficients.} Computed from 45-qubit simulations of two coupled 22-qubit 1D chains joined by a 2D interaction rung. The plots show $|T_{\text{same}}|$, $|T_{\text{cross}}|$, and $|R|$ as functions of time steps $N$ for three interaction cases: uncoupled 1D, $\vec{\lambda} = (1,1,0)$, and $\vec{\lambda} = (0,0,1)$ with a fixed interaction strength ratio $J_{\perp}/J = 4$. Only $T_{\text{cross}}$ transmission is observed for $\vec{\lambda} = (1,1,0)$, indicating reduced reflection and enhanced transport resilience.
\textbf{(b) Directional dependence of superdiffusion breakdown.} Heatmaps show how breakdown varies with the correlation measurement direction $\hat{n} = n_x \hat{x} + n_z \hat{z}$ for $\vec{\lambda} = (0,0,1)$ and $\vec{\lambda} = (1,1,0)$, with fixed interaction strength ratio $J_{\perp}/J = 1$. $\vec{\lambda} = (0,0,1)$ transitions to ballistic or diffusive behavior depending on $\hat{n}$, while $\vec{\lambda} = (1,1,0)$ remains resilient across all directions.  Dashed lines serve as guides to the eye. 
}
    \label{scattering_combined}
\end{figure*}

\subsection{Noiseless simulations}

We simulate a 28-qubit system based on the model described in Section~\ref{model_description} to study the breakdown of superdiffusion under different 2D interaction types and the relative position of the 2D interaction rung to the probe qubit, as shown in Fig.~\ref{interaction_type_influence}. All simulations use a fixed interaction strength ratio of $\frac{J_{\perp}}{J} = 1.0$. The figure shows running averages of scaling exponents $\phi(t) = \left\langle \frac{d \ln(C^{zz}(t))}{d \ln(N)} \right\rangle$ as a function of time steps $N$. Additional results are provided in Appendix~\ref{correl_values}, including simulations with two 2D rungs using a 29-qubit system and varying interaction strength ratios $\frac{J_{\perp}}{J}$ from $10^{-4}$ to 4, along with corresponding correlation values.

\subsubsection*{Effect of different interaction types}

Our experiments in Fig.~\ref{interaction_type_influence} demonstrate that the nature of the 2D interaction type significantly influences the breakdown of superdiffusion. We observe deviations in the scaling exponent from the superdiffusive regime ($-2/3$) toward either the diffusive ($-1/2$) or ballistic ($-1$) regimes, with varying degrees of resilience across interaction types.

Interaction types $\vec{\lambda} = (0,0,1)$ and $\vec{\lambda} = (1,1,0)$ preserve the conservation of total spin $S = \sum_i \frac{\sigma_z^i}{2}$, parity $P = \prod_i \sigma_x^i$, and the existence of an odd current $j_x P = -P j_x$. These symmetries imply the absence of ballistic transport channels based on locally conserved charges~\cite{ballistic_absent}. Accordingly, these interaction types exhibit breakdown toward the diffusive regime, as shown in Fig.~\ref{interaction_type_influence}, with $\vec{\lambda} = (1,1,0)$ showing greater resilience than $\vec{\lambda} = (0,0,1)$.

Interaction types $\vec{\lambda} = (1,0,0)$ and $\vec{\lambda} = (1,0,1)$ conserve parity but break total spin conservation. Both show breakdown toward the ballistic regime, with $\vec{\lambda} = (1,0,1)$ exhibiting higher resilience than $\vec{\lambda} = (1,0,0)$. 

The interaction type $\vec{\lambda} = (1,1,1)$ retains total spin conservation, parity, and the full $SU(2)$ symmetry of the 1D Heisenberg model. It shows the highest resilience against superdiffusion breakdown among all interaction types. In the Appendix \ref{correl_values}, we see that they require 2D interaction strengths of $J_\perp \sim 4J$  to exhibit any deviation toward the diffusive regime. This observation aligns with previous findings on $SU(2)$-preserving perturbations in spin ladder systems~\cite{breaking_quantum}. 

We emphasize that we have used the terms "diffusive breakdown" and "ballistic breakdown" only to describe the direction of deviation in the scaling exponent, rather than convergence to exact transport coefficients. A shift from $-2/3$ toward $-1/2$ indicates diffusive breakdown, while a shift toward $-1$ indicates ballistic. In the ballistic case, where $z$-spin is not conserved, spin transport becomes ill-defined.

\subsubsection*{Position of the probe qubit relative to the 2D rungs}

The position of the 2D interaction rung dictates the onset of the superdiffusion breakdown and doesn't affect the nature and relative extents of the breaking. This is elucidated by the interaction type $\vec{\lambda} = (0,0,1)$ showing a clear delayed onset of the superdiffusion breakdown when the only active 2D interaction is the farther rung in Fig \ref{interaction_type_influence}. This delayed onset of the superdiffusion breakdown could be attributed to the longer time taken by the 2D interaction to impact the probe qubit during the time evolution. And we show in Appendix \ref{correl_values} that turning 2D interactions on both rungs shows a nearly cumulative effect of individual rungs.

\subsubsection*{Relative Resilience against the breakdown}

To explain the relative resilience against superdiffusion breakdown,  
for interaction types where spin transport remains well-defined, we compute scattering coefficients using 45-qubit simulations consisting of two coupled 22-qubit 1D chains joined by a 2D interaction rung (see Fig.~\ref{scattering_combined}(a)). Matrix product state simulations of the Trotterized circuit were performed with a maximum bond dimension of $\chi = 512$. These configurations closely resemble our model, as finite-size effects from chain boundaries have not yet emerged at the simulated timescales. Reflection and transmission coefficients are obtained by summing the correlators $C^{zz}_{ip}(t)$ over relevant sites and taking the absolute value, where $p$ denotes the probe qubit. Here, $R$ represents reflection (i.e., spin remaining before the rung), $T_{\text{same}}$ is transmission along the same chain, and $T_{\text{cross}}$ is transmission to the other chain. 

Notably, in the $\vec{\lambda} = (0,0,1)$ (equivalently, ZZ) interaction case, no $T_{\text{cross}}$ transmission is observed (i.e., spin is not transported across the rung). As a result, the reflection channel $R$ is enhanced. The autocorrelator effectively measures the spin that has reflected back to the initial site, so one therefore sees a reduced resilience. One can understand this lack of cross-chain transmission by considering the Heisenberg evolution of the probe qubit, $\sigma_p^z(t)$. Conjugating by any $e^{-ih_{i,j}\tau}$ preserves the form
\[
\sigma_p^z(t) = A_{\text{chain-1}}(t) \otimes I_{\text{rung}} \otimes I_{\text{chain-2}} + B_{\text{chain-1}}(t) \otimes \sigma^z_{\text{rung}} \otimes I_{\text{chain-2}}
\]
for the given rung interaction, so the operator can be nontrivial on the rung site but not past it. On the other hand, in the $\vec{\lambda} = (1,1,0)$ (equivalently, XX+YY) interaction case, $T_{\text{cross}}$ transmission is observed, indicating reduced reflection and thus greater resilience.

The resilience observed across different interaction types is not limited to a specific direction of correlation measurement. As shown in Fig.~\ref{scattering_combined}(b), interaction types other than $\vec{\lambda} = (1,1,1)$ can exhibit distinct breakdown behavior depending on the spin direction along which correlations are measured. Notably, the less resilient interaction type $\vec{\lambda} = (0,0,1)$ shows deviation toward ballistic scaling when correlations are measured along the $x$ direction instead of the $z$ direction. In contrast, the more resilient interaction type $\vec{\lambda} = (1,1,0)$ does not exhibit such sensitivity and remains resilient against breakdown across all measurement directions.

\begin{figure}[ht!]
   
    \centering
\includegraphics[width=1\columnwidth]{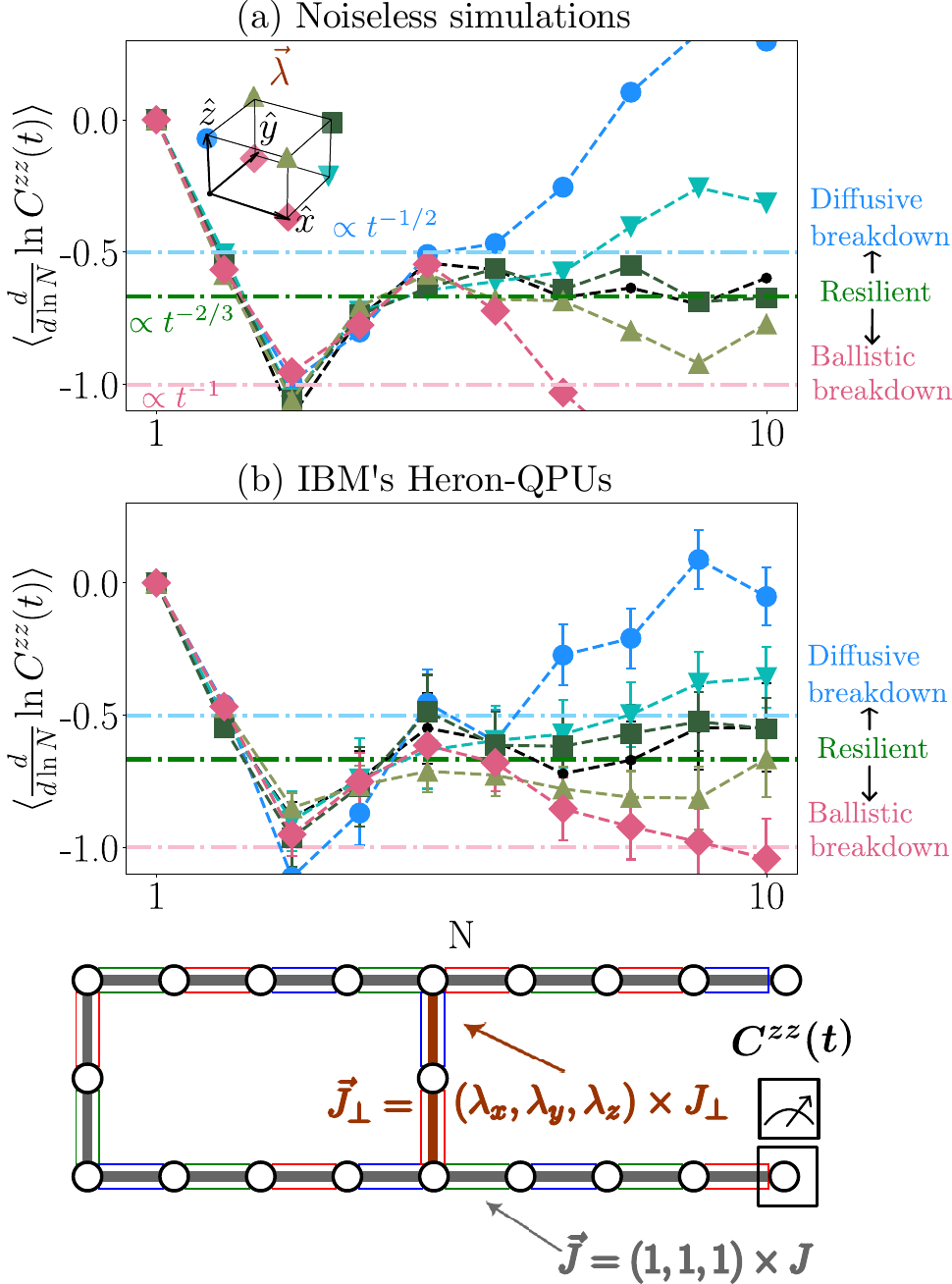}
        \caption{ \textbf{Comparison of superdiffusion breakdown in noiseless simulations (a) and IBM Heron QPU experiments (b) for various 2D interaction types:} The plots display running averages of scaling exponents with propagated standard errors derived from correlation measurements. Early-time agreement validates the simulation protocol, while intermediate-time deviations demonstrate the QPU’s capability to resolve the relative resilience of interaction types against superdiffusion breakdown, despite noise introduced by deep circuits involved in the protocol.}
     \label{hardware_runs}
\end{figure}

\subsection{Hardware simulations}
 Our noiseless simulations indicate that the breakdown of superdiffusion is primarily governed by the type of 2D interaction. In this section, we investigate the feasibility of observing the influence of interaction types on superdiffusion breakdown by simulating a minimal system consisting of 20 qubits from the model shown in Fig. \ref{fig:Scheme}, using IBM's Heron Quantum Processing Units (QPUs).
 
The depth of circuits in the hardware simulations dictates the noise susceptibility of our results. This comes from the trotterization and the random state preparation involved the algorithm \cite{Richter_Pal}. Depth from the former can be reduced by keeping the overall time of evolution fixed but choosing higher $\tau$. However, choosing a $\tau >> 1$ would take us far away from the physically relevant continuous time model. In Appendix \ref{tau_optimisation}, we show that $\tau = 1$ captures the superdiffusion breakdown features of the continuous time model, keeping the overall number of time steps, and hence the depth, required under control. The depth from random state preparation can be reduced by picking the optimal number of cycles and averaging over multiple runs to approximate the results from a true random state or random state prepared with a high number of cycles. See Appendix \ref{randomization_section} for details. These optimizations resulted in arriving at the favorable parameters $\tau = 1$ and $c = 9$ cycles used for random state preparation. The results were averaged over five runs each containing $2 \times 10^4$ shots. Further, error mitigation techniques like Dynamical Decoupling \cite{DD,pokharel2018demonstration} was used in the final report of our results.

Figure \ref{hardware_runs}(b) shows the superdiffusion breakdown observed in IBM’s Heron processors (see Appendix \ref{device_specs} for calibration details) for different interaction types, along with a comparison to noiseless simulations shown in panel (a), run with the same parameter settings ($c$, $\tau$). The running averages of the scaling exponents in the plots show remarkable agreement with the noiseless simulations at early times. At intermediate times, although the extent of the breakdown (i.e., deviation from the superdiffusive scaling exponent) is not captured as accurately as in earlier times, the relative resilience of the interaction types against superdiffusion breakdown is clearly observed. The error bars in the plot are obtained through propagation of standard errors from individual correlation measurements (see Appendix \ref{prop_std} for details). We also note that the interaction strength used in the hardware experiments was chosen to be higher than in the noiseless simulations shown in Fig. \ref{interaction_type_influence}, with $\frac{J_{\perp}}{J} = 4.0$. This choice ensures that the breakdown occurs around intermediate times rather than later, since longer circuit depths would make the hardware significantly more susceptible to noise, limiting its usefulness in identifying superdiffusion breakdown and the resilience of different interaction types. 

These results could be further improved by compressing high circuit depths during late-time evolution using various circuit compression techniques such as those in \cite{AQC}, and advanced error mitigation methods like Probabilistic Error Cancellation and Amplification (PEC/PEA) \cite{EM_4}. However, since the correlation values and the probabilities of capturing relevant bitstrings are on the order of $O(0.01)$, these methods may require significantly more time to learn the noise model due to the increased accuracy demands.

\section{Conclusion}

To connect with real materials, which often include imperfections and integrability-breaking terms, we extended a one-dimensional integrable superdiffusive model to a two-dimensional version compatible with heavy-hex lattice geometry and tunable interactions. Our quantum simulations reveal that superdiffusion breaks down with varying resilience toward either diffusive or ballistic regimes, depending on the interaction type. We categorized interaction types based on their relative resilience, suggesting a pathway to sustaining superdiffusive spin transport in two-dimensional lattices. 

We also provided insights into the mechanisms that influence this resilience, including directional scattering effects and anisotropic transport behavior that merit deeper investigation. 

Crucially, the model is natively implementable on current quantum hardware, and the observed breakdown behavior was reproduced on IBM’s Heron QPUs with remarkable accuracy. With improved error mitigation~\cite{EM_review, EM_review2, EM_review3} and access to more flexible hardware topologies, this framework can be scaled to larger systems and experimentally relevant material lattices. This opens new avenues for probing non-equilibrium quantum dynamics in regimes beyond classical reach and for connecting quantum simulations to quasi-one-dimensional antiferromagnets such as KCuF$_3$ and CsCoCl$_3$, where weak interchain couplings and anisotropies introduce deviations from ideal integrable behavior.

\section{Acknowledgements}
We acknowledge funding from the Office of Science through the Quantum Science Center (QSC), a National Quantum Information Science Research Center. We acknowledge the use of IBM Quantum services for this work. We also acknowledge Joel Moore, Norhan Eassa, Varadharajan Muruganadham, and Zoe Holmes for insightful discussions.

%%%%%%%%%%%%%%%%%%%%
%%% BIBLIOGRAPHY %%%
%%%%%%%%%%%%%%%%%%%%
\bibliography{ref}

\widetext
\pagebreak
%\newpage
\appendix

\AtAppendix{\counterwithin{lemma}{section}}

\section{Quantum Algorithm: Implementation details}
\label{implementation_details}
The infinite temperature $ZZ$ spin spin auto-correlation function on site $i$ is given by 
\begin{equation}
\label{2_point}
C^{zz}_{ii}(t) = \text{Tr} (\sigma_z^i(0)\sigma_z^i(t))/ 2^n
\end{equation}
 where $n$ is the number of spins in the system. We follow the algorithm proposed by Richter and Pal \cite{Richter_Pal} in computing this quantity. This algorithm involves an initial state preparation step where a Haar random state $\ket {\Psi_R}$ is prepared with support on all qubits except the $i$-th qubit so that the initial state is $\ket{\Psi_{R,i}} = \ket{0}_i\ket{\Psi_R}$. With this state, the correlation function in Eq \ref{2_point} can be written as a single point observable as in Eq \ref{1_point}. 
 
 \begin{equation}
     \label{1_point}
     C_{ii}(t) = \frac{1}{2} \langle \Psi_{R,i} | \sigma^i_z(t) | \Psi_{R,i} \rangle + O(2^{-n/2})
 \end{equation}
We prepare $\ket {\Psi_R}$ by performing $c$ cycles of applying random two-qubit unitaries on the $r$, $g$, and $b$-type bonds (Fig \ref{fig:Scheme}) in that order.  The number of cycles  ($c$) required for reliable inference of the superdiffusion breakdown is discussed in Section \ref{randomization_section}. 

 \subsection{Trotterization}

 To compute the expectation of $\sigma_z^i(t)$ in  Eq \ref{1_point}, we evolve the initial state $\ket{\Psi_{R,i}}$ as described in Eq \ref{2D_evol}. This involves the application of two-qubit gates at $r$, $g$, and $b$-type bonds (Fig \ref{fig:Scheme}) for the two local Hamiltonian terms acting on them.\\
 
 If a two-local Hamiltonain, $h$,  acts on sites $k$ and $l$ with interaction given by 
 $J_h (\frac{\lambda_x}{4} \sigma^k_x \sigma^l_x + \frac{\lambda_y}{4} \sigma^k_y \sigma^l_y  +\frac{\lambda_z}{4} \sigma^k_z \sigma^l_z )$,  one realisation of the trotter circuit corresponding to the time evolution of $h$ is given by Fig \ref{trotter_unit}.

 \begin{figure}[htbp!]
    \centering
    \begin{tikzpicture}
        \node[scale=1.0]{\begin{quantikz} 
     & \ctrl{1} &\gate[style={fill=orange!20}]{RX(\frac{J_h \lambda_x t}{2} )}&\gate[style={fill=purple!60}]{H} & \ctrl{1} &\gate[style={fill=orange!60}]{P (-\frac{\pi}{2})}&\gate[style={fill=purple!60}]{H}& \ctrl{1} & \gate[style={fill=orange!20}]{RX(\frac{\pi}{2})}&  \\ 
     & \targ{}&\gate[style={fill=purple!20}]{RZ(\frac{J_h \lambda_z t}{2} )}&&\targ{}&\gate[style={fill=purple!20}]{RZ(-\frac{J_h \lambda_y t}{2} )}&&\targ{}&\gate[style={fill=orange!20}]{RX(-\frac{\pi}{2})}&\\
     \end{quantikz}};
    \end{tikzpicture}
    \caption{\textbf{Trotter circuit for the time evolution of $J_h \times \vec{\lambda} = J_h (\lambda_x,\lambda_y,\lambda_z)$ for time $t$:} The circuit is equal to $e^{-i J_h (\frac{\lambda_x}{4} \sigma_x \sigma_x + \frac{\lambda_y}{4} \sigma_y \sigma_y  +\frac{\lambda_z}{4} \sigma_z \sigma_z ) t }$ up to a global phase.}
    \label{trotter_unit}
\end{figure}

\subsection{$\tau$ dependence}

\begin{figure*}
    \centering
    \includegraphics[width=0.85\textwidth]{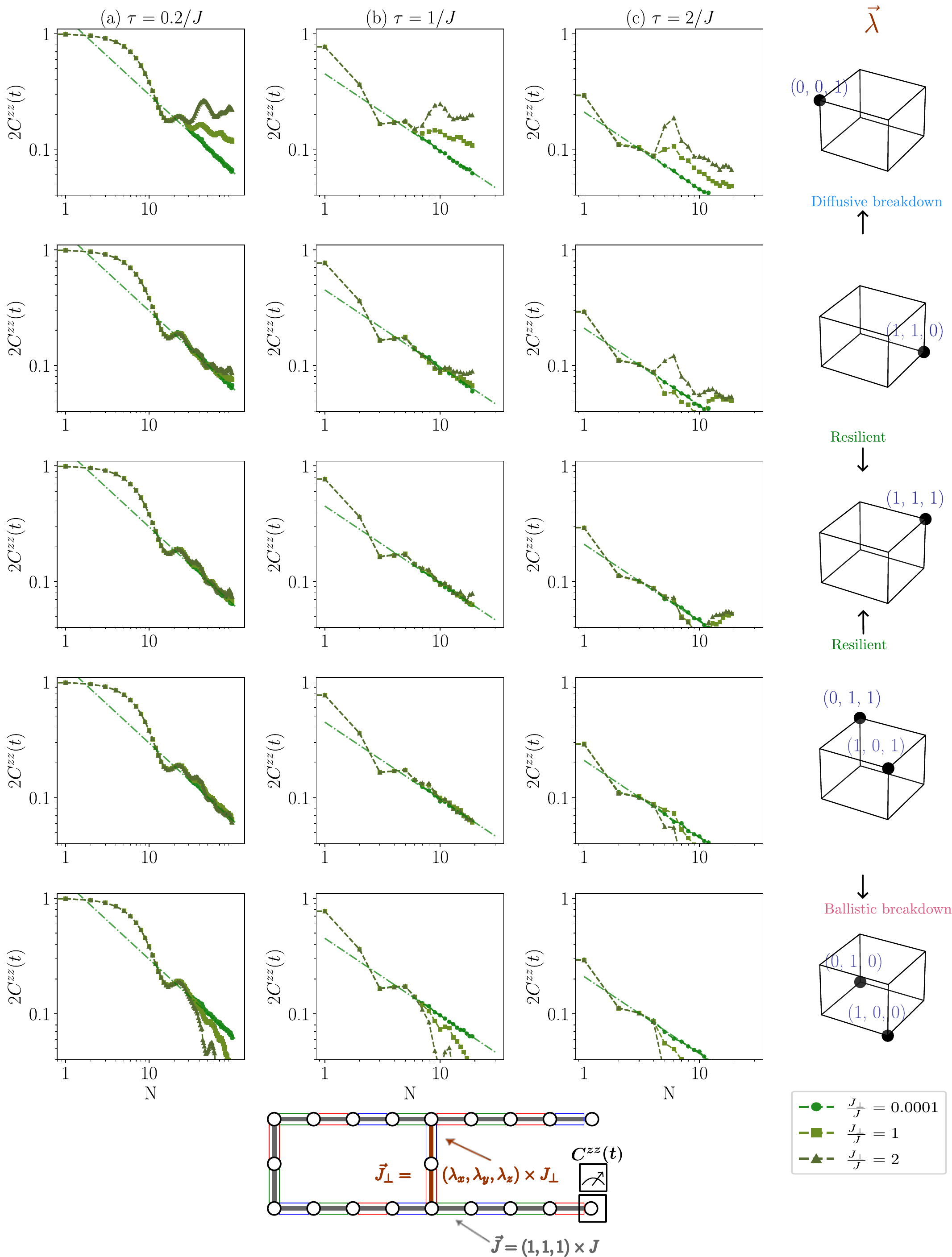}
    \caption{\textbf{$\tau$ optimisation}: We study the effect of $\tau$ on the superdiffusion breakdown in our model. We do so by setting $J = 1.0$ and tuning up the $\tau$.  Dash-dotted lines in each plot indicate the superdiffusive decay. (a) We reduce the trotter error by setting a value of $\tau$ as low as 0.2 to effectively mimic the continuous time model. We observe the superdiffusion breakdown to diffusion and ballistic regime happening even at this limit. We also observe the relative resilience among interaction types as discussed in the main text. (b) By tuning up $\tau$ to 1.0, we reproduce the features of the continuous time model but using short depth (fewer trotter steps), hence hardware-friendly, quantum circuits. This value of $\tau$ was used in our noiseless and hardware simulations. (c) Tuning up $\tau$ to 2.0 quickens the equilibration process and reaches the tail end of evolution quicker. Hence, the time evolution sometimes does not capture the features that we observe in (a), see $\lambda = (1,1,1)$ for example.  Moreover, setting higher $\tau$ takes our discrete model far away from the continuous model and we do not expect it to show the features we observe in (a).}
    \label{dt_optimisation}
\end{figure*}

The model described in section \ref{model_description}
comes with an inherent parameter $\tau$. While setting a low value of $\tau << 1$ mimics the behavior of a continuous time model, it requires a higher number of trotter steps (and hence depth) to evolve till a certain final time $t$ and will be significantly susceptible to noise when implemented in quantum hardware. However, considering $\tau >> 1$ takes our discrete model far away from the continuous model and loses any physical relevance. In Fig \ref{dt_optimisation}, we show that (b) $\tau = 1$ can still capture the superdiffusion breakdown that exists in the continuous model (a)  $\tau = 0.2$ with fewer trotter steps. Choosing higher $\tau$ results in quicker equilibration and we fail to see the superdiffusion breakdown clearly. 
\label{tau_optimisation}

\newpage
\section{Correlation values $C^{zz}(t)$ of the experiments}
\label{correl_values}
\begin{figure*}[h]
    \centering
    \includegraphics[width=0.8\textwidth]{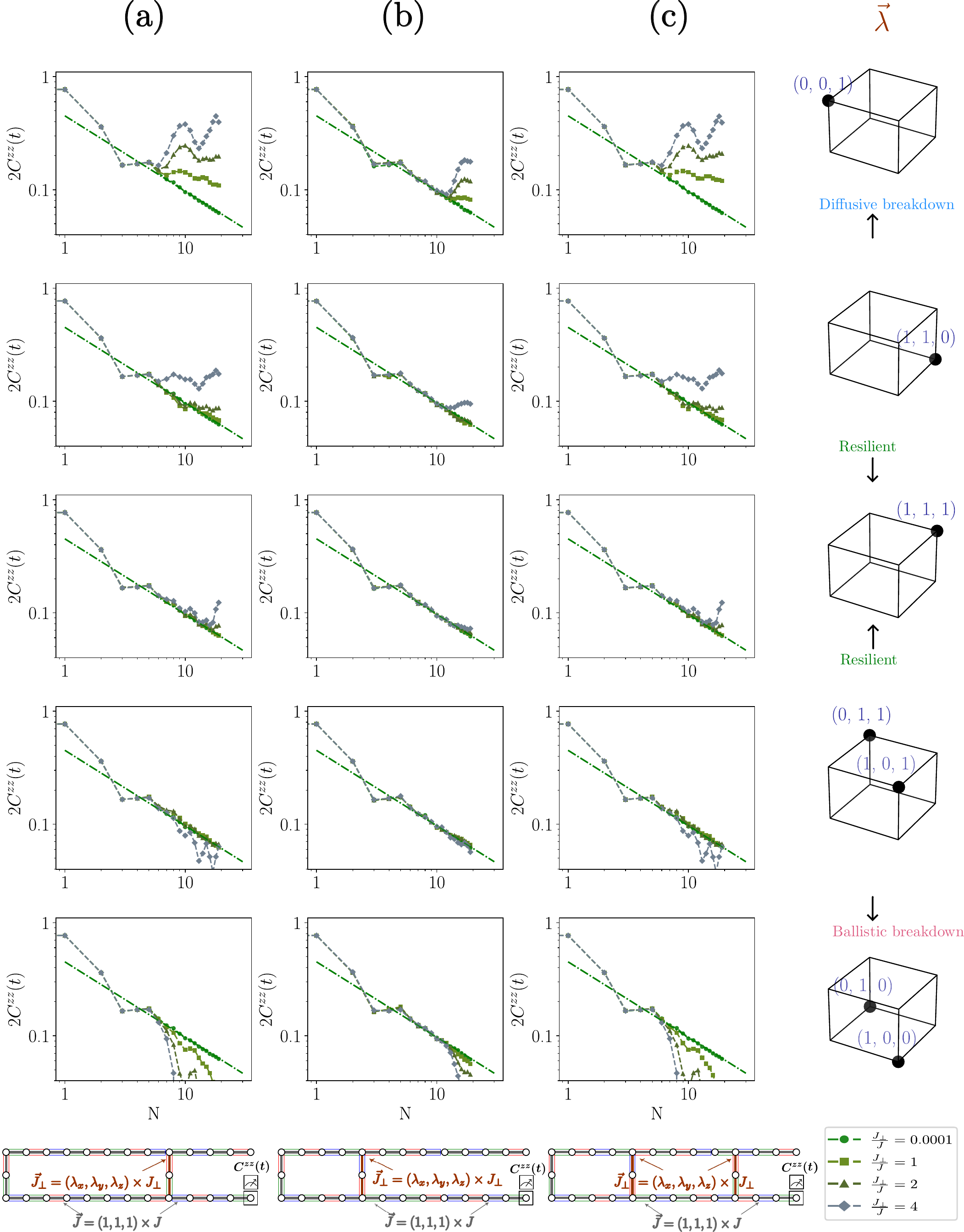}
    \caption{Correlation values $C^{zz}(t)$ for 2D interaction strengths from $10^{-4}$ to $4.0$, illustrating superdiffusion breakdown through qualitative deviations from the power law decay. Dash-dotted lines in each plot indicate the superdiffusive decay.}
    \label{29_qubits_correl_values}
\end{figure*}

\begin{figure*}
    \centering
    \includegraphics[width=0.8\textwidth]{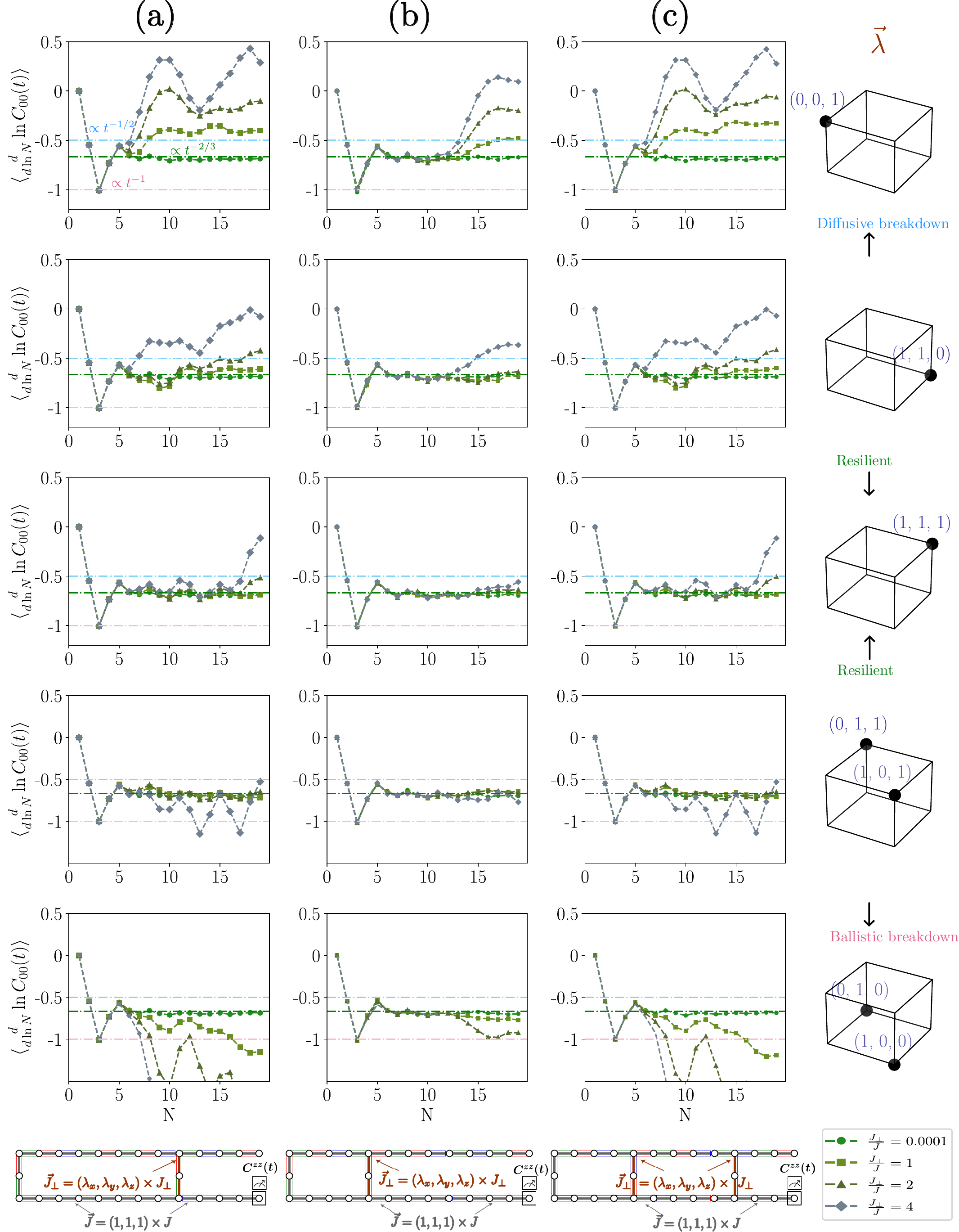}
    \caption{Scaling exponents for 2D interaction strengths from $10^{-4}$ to $4.0$, illustrating superdiffusion breakdown through quantitative deviations from the power law decay. 
}
    \label{29_qubits_ra_values}
\end{figure*}

The correlation values ($C^{zz}(t)$) (Fig \ref{29_qubits_correl_values}) and the corresponding running averages (Fig \ref{29_qubits_ra_values}) of scaling exponents are displayed in this section for 2D interaction strengths ranging from $10^{-4}$ to $4.0$. While the scaling exponents elucidate the superdiffusion breakdown quantitatively, we see that the plots containing correlation values elucidate the same qualitatively. The deviation from the linear superdiffusive decay towards both diffusive and ballistic sides is captured in these plots. 

\newpage

\section{Number of cycles ($c$) used in the random state preparation}
\label{randomization_section}
 \begin{figure*}
    \centering
    \includegraphics[width=1\textwidth]{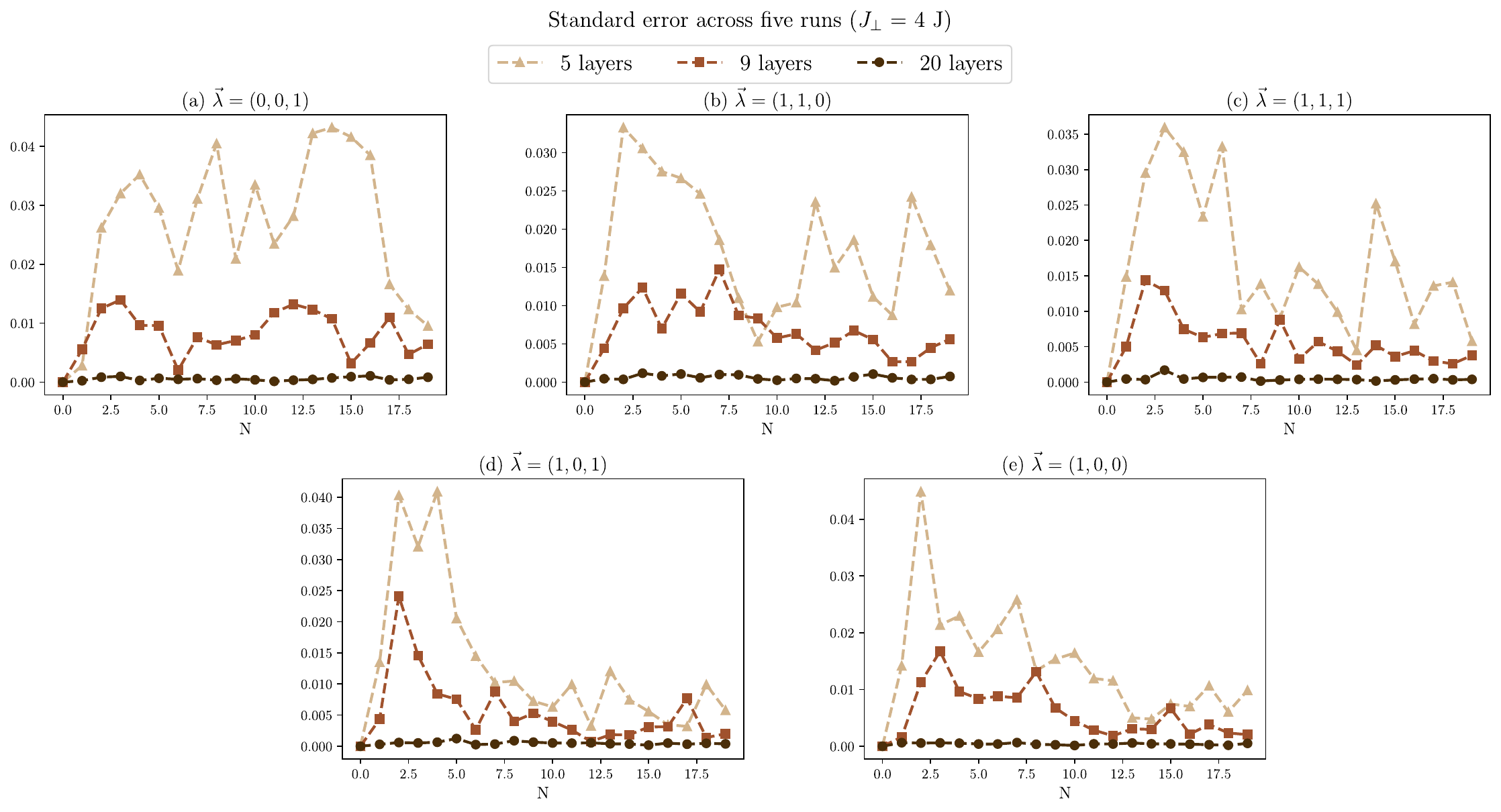}
    \caption{\textbf{Standard error of correlation values across five runs:} The figure shows the standard error in correlation values, computed as the mean over five independent runs for different parameter settings ($c$) in the 20-qubit system described in Section~\ref{model_description}. The standard deviation is normalized by the square root of the number of runs (five). As the number of cycles increases from 5 to 9 and 20, the standard deviation decreases from $\geq 0.01$ at late times to $\sim (0.001)$.}
    \label{averaging}
\end{figure*}
Fig \ref{averaging} shows the standard error in correlation values computed as a mean over five independent runs for different number of cycles $c$ used in the random state preparation. $c = 20$ has a very low standard deviation in the $\sim  0.001$ during late times and has high resolution for inferring the superdiffusion breakdown, as the correlation values are known to drop to only $\sim $ 0.01 during late times. However, $c=20$ cycles implementation in hardware requires higher depth and might be susceptible to noise. So we look at lower $c$ values and we observe that  $c=9$ shows a standard deviation of around $ \sim  0.005$ during late times which is already good enough for capturing the superdiffusion breakdown and requires relatively short depth compared to 20 layers. If we further lower the $c$ value to 5, the standard deviation is almost close to $\geq 0.01$ and hence is not suitable for capturing superdiffusion breakdown. This led us to our choice of using $c=9$ in our hardware implementation Fig \ref{hardware_runs}.

\section{Propogation of standard errors}
\label{prop_std}
Scaling exponent at the $i$-th step is defined as

\begin{equation}
Y_i = \frac{\log(C_{i+1}) - \log(C_i)}{\log(t_{i+1}) - \log(t_i)},
\end{equation}

where \( C_i \) denotes the correlation value at time \( t_i \). The uncertainty in each \( Y_i \) is obtained via standard error propagation:

\begin{equation}
\sigma_{Y_i} = \frac{1}{\log(t_{i+1}) - \log(t_i)} \cdot \sqrt{ \left( \frac{\sigma_{C_{i+1}}}{C_{i+1}} \right)^2 + \left( \frac{\sigma_{C_i}}{C_i} \right)^2 },
\end{equation}

assuming independent uncertainties in \( C_i \) and \( C_{i+1} \).

To compute the uncertainty in the running average of scaling exponents up to  \( i \)-th step, we define

\begin{equation}
\bar{Y}_i = \frac{1}{i+1} \sum_{j=0}^{i} Y_j,
\end{equation}

and propagate the errors using:

\begin{equation}
\sigma_{\bar{Y}_i} = \frac{1}{i+1} \sqrt{ \sum_{j=0}^{i} \sigma_{Y_j}^2 }.
\end{equation}

This expression assumes that the individual slope estimates \( Y_j \) are statistically independent. The resulting standard errors \( \sigma_{\bar{Y}_i} \) are plotted as error bars in the Fig \ref{hardware_runs}.

\section{Device specifications}
\label{device_specs}

We performed hardware runs on IBM's Heron processors, specifically \textit{ibm\_marrakesh} and \textit{ibm\_pittsburgh}. We report the calibration data of these processors at the time of our runs below. 
$T_1$ ($\mu$s), $T_2$ ($\mu$s), and CZ errors were computed for the subset of qubits used in our runs. 
Readout error was computed for the probe qubit only.

\begin{itemize}
    \item On \textit{ibm\_pittsburgh}, we utilized a linear qubit chain comprising qubits \texttt{61, 62, 63, \ldots, 69, 78, 89, 88, 87, \ldots, 81}. The 2D interaction rung qubit was qubit \texttt{77}, and the probe qubit was qubit \texttt{61}.
    \begin{table}[h!]
\centering
\caption{\textit{ibm\_pittsburgh} Calibration Data Summary}
\begin{tabular}{|l|c|c|c|}
\hline
\textbf{Calibration Data} & \textbf{Min} & \textbf{(Mean, Median)} & \textbf{Max} \\
\hline
$T_1$ ($\mu$s) & 132.420 & (314.655, 319.722) & 443.791 \\
$T_2$ ($\mu$s) & 177.944 & (365.463, 346.771) & 529.992 \\
Readout Error & 0.002197 & (0.002197,0.002197) & 0.002197 \\
CZ Error & 0.000983  & (0.001615, 0.001400) & 0.003508\\
\hline
\end{tabular}
\label{tab:calibration_data_pittsburgh}
\end{table}

    \item On {\textit{ibm\_marrakesh}}, the linear chain included qubits \texttt{43, 44, \ldots, 51, 58, 71, 69, 68, \ldots, 63}. The 2D interaction rung qubit was qubit \texttt{57}, and the probe qubit was qubit \texttt{43}.
\begin{table}[h!]
\centering
\caption{{\textit{ibm\_marrakesh}} Calibration Data Summary}
\begin{tabular}{|l|c|c|c|}
\hline
\textbf{Calibration Data} & \textbf{Min} & \textbf{(Mean, Median)} & \textbf{Max} \\
\hline
$T_1$ ($\mu$s) & 116.910 & (199.429, 192.971) & 288.766 \\
$T_2$ ($\mu$s) & 22.157 & (138.292, 123.492) & 358.621 \\
Readout Error &  0.009766 & (0.009766, 0.009766) & 0.009766 \\
CZ Error & 0.002177 & (0.003632, 0.0031432) & 0.007512 \\
\hline
\end{tabular}
\label{tab:calibration_data_marrakesh}
\end{table}

\end{itemize}

\end{document}